 \documentclass{article}

\usepackage{epsfig} 
\usepackage{amsfonts}
\usepackage{graphicx}

 \usepackage{amssymb}

\date{\today}

\newcommand{\insertplot}[5]{\begin{figure}
 \hfill\hbox to 0.05in{\vbox to #5in{\vfill
 \inputplot{#1}{#4}{#5}}\hfill}
 \hfill\vspace{-.1in}
 \caption{#2}\label{#3}
 \end{figure}}
 \newcommand{\inputplot}[3]{
 \special{ps: plotfile #1}
}
\newcounter{fig}

\newcommand{\ee}{\end{equation}}
\newcommand{\eea}{\end{eqnarray}}
\newcommand{\be}{\begin{equation}}
\newcommand{\bea}{\begin{eqnarray}}

\begin{document}

\title{
Myers-Perry black holes with scalar hair and a mass gap: \\ unequal spins 
 }

\author{
{\large Carlos Herdeiro}$^{1}$,
{\large Jutta Kunz}$^{2}$,
{\large Eugen Radu}$^{1}$
 and
{\large Bintoro Subagyo}$^{3}$
\vspace{0.5cm}
\\
$^{1}${\small Departamento de F\'\i sica da Universidade de Aveiro and CIDMA, Campus de Santiago, 3810-183 Aveiro, Portugal}
   \\
$^{2}${\small Institut f\"ur  Physik, Universit\"at Oldenburg, Postfach 2503 
  D-26111 Oldenburg, Germany}
   \\
$^{3}${\small Department of Physics, Institut Teknologi Sepuluh Nopember, Indonesia
 }
} 
\date{May 2015}
  \maketitle 
 
\begin{abstract} 
We construct rotating boson stars and Myers-Perry black holes 
with scalar hair (MPBHsSH) as fully non-linear solutions of  five dimensional Einstein gravity minimally coupled to a complex, massive scalar field.
 The MPBHsSH are, in general, regular on and outside the horizon, asymptotically flat, and possess angular momentum in a single rotation plane. 
 They are supported by rotation and have no static limit. Such hairy BHs may be thought of as bound states of boson stars and singly spinning, vacuum MPBHs and inherit properties of both these building blocks. When the horizon area shrinks to zero, the solutions reduce to (in a single plane) rotating boson stars; but the extremal limit also yields a zero area horizon, as for singly spinning MPBHs. Similarly to the case of equal angular momenta, and in contrast to Kerr black holes with scalar hair, singly spinning MPBHsSH are disconnected from the vacuum black holes, due to a mass gap. We observe that for the general case, with two unequal angular momenta,  the equilibrium condition for the existence of MPBHsSH is $w=m_1\Omega_1+m_2\Omega_2$, where $\Omega_i$ are the horizon angular velocities in the two independent rotation planes and $w,m_i$, $i=1,2$, are the scalar field's frequency and azimuthal harmonic indices. 
\end{abstract}



\section{Introduction and motivation}
Apart from vacuum and electro-vacuum, scalar-vacuum is the simplest model that may be considered in Einstein gravity. In its simplest form, this theory corresponds to couple (minimally) to gravity one or more real massless scalar fields with standard kinetic terms and without self-interactions. Unlike electro-vacuum, however, such scalar-vacuum does not yield any new stationary, asymptotically flat and regular black hole (BH) solutions, as compared to pure vacuum. This conclusion is based on a four dimensional \textit{no-scalar-hair} theorem~\cite{Chase:1970} 
(see~\cite{Herdeiro:2015waa} for a review). The physics rationale is twofold. Firstly, scalar fields do not have an associated Gauss law, albeit they may have a local conservation law, for instance, if there is a global symmetry. Thus, if some amount of scalar field falls into a BH, then, at least classically, no memory of it is expected to be found in the exterior spacetime. Secondly, some amount of a free, minimally coupled scalar energy placed in the neighborhood  of a BH is expected to either disperse to infinity or be absorbed by the BH. And neither of these fates endows the BH spacetime with an eternally lingering scalar field in the vicinity of the event horizon.

\bigskip

A minimal addition to scalar-vacuum, however, produces a remarkable change of affairs. Adding a mass term in a theory with two equally massive real scalar fields, or equivalently, with a single massive complex scalar field, new regular, asymptotically flat BH solutions exist, both in four spacetime dimensions ($D=4$)  -- Kerr BHs with scalar hair~\cite{Herdeiro:2014goa,Herdeiro:2014ima,Herdeiro:2015gia} -- and in $D=5$ -- Myers-Perry BHs with scalar hair (MPBHsSH)~\cite{Brihaye:2014nba} 
 (see also the recent work~\cite{Kleihaus:2015iea} 
for a $D=4$  generalization). 
The underlying physics justifying the existence of non-trivial scalar fields in these two examples has clear differences. In the Kerr case, the new solutions can be inferred at the linear level due to the existence of test field scalar clouds at the threshold of superradiant instabilities~\cite{Hod:2012px,Hod:2013zza,Herdeiro:2014goa}. In the Myers-Perry case, by contrast, there are no superradiant instabilities for a massive scalar field~\cite{Cardoso:2005vk}; the scalar hair found in~\cite{Brihaye:2014nba} is intrinsically non-linear and  originates a mass gap between the hairy and the vacuum Myers-Perry BHs. But similarities exist: in both cases $i)$ the gravitational theory admits asymptotically flat, everywhere regular, solitonic solutions without a horizon, \textit{boson stars}~\cite{Schunck:2003kk,Liebling:2012fv}, for which the scalar field has a harmonic time dependence with frequency $w$; $ii)$ the hairy BH solutions can be regarded as adding a rotating BH horizon within a spinning boson star~\cite{Herdeiro:2014ima}, with the hairy BHs inheriting properties of both these building blocks;
in particular whereas the $D=4$ boson stars continuously connect to Minkowski spacetime, the $D=5$ boson stars (with two equal rotations) already possess a mass gap with respect to the Minkowski vacuum~\cite{Hartmann:2010pm}; $iii)$ a central condition for the existence of  all known scalar 
hairy BHs relies on the identification of the horizon null generator with the Killing vector field that preserves the rotating boson star solution~\cite{Dias:2011at}.

\bigskip

The $D=5$ case studied in Ref.~\cite{Brihaye:2014nba} pertained solutions with two complex scalar fields and two equal angular momenta parameters, as this choice leads to a co-dimension one problem and thus considerable technical simplification. 
The corresponding $D=5$ Myers-Perry BHs are akin to the $D=4$ Kerr solution; 
in particular they are both a two parameter family of solutions -- characterized, say, by the ADM mass, $M$, and horizon angular velocity, $\Omega_H$ -- and have a regular, finite area, extremal limit. In both cases the hairy BHs, just as the boson stars, have (a) monochromatic scalar field(s) whose frequency $w$ is fixed by $\Omega_H$ and (in $D=4$) an azimuthal winding number.

\bigskip

The single angular momentum $D=5$ Myers-Perry solution, by contrast, is singular in the extremal limit, 
 while the generic solution with two angular momenta is characterized by two different horizon angular velocities $\Omega_1,\Omega_2$.
 We would therefore like to understand if this more general case can still accommodate scalar hair 
 and if so how the scalar field frequency relates the two angular velocities.
  In this paper we shall clarify both these issues. 
  We show that the equilibrium condition for the general case with two non-vanishing angular momenta is:
\begin{equation}
w=m_1\Omega_1+m_2\Omega_2 \ ,
\label{equilibrium}
\end{equation}
where $m_i$ are the two azimuthal quantum numbers in the scalar field ansatz, $cf.$ eq.~(\ref{scalar}) below. Actually to reach this conclusion it is not necessary to solve the fully non-linear systems, as condition (\ref{equilibrium}) can be derived from regularity at the horizon. We will then focus our analysis of the fully non-linear system on the case with a single angular momentum parameter, and we shall derive both the corresponding boson star solutions and the hairy BHs. The former solutions again show the property of their cousins with two equal angular momenta in \cite{Brihaye:2014nba}: 
they do not trivialize in the limit of maximal allowed frequency 
and exhibit a mass gap with respect to Minkowski spacetime.
 The latter solutions have a domain of existence delimited, in particular, by extremal solutions which are singular,
  in agreement with the behaviour of the hairless singly spinning Myers-Perry BHs. 
This reinforces the picture of these hairy BHs as {\it ``horizons inside classical lumps"},
 the classical lumps being boson stars in this case, wherein such a bound state inherits properties 
 of both the solitonic limit and of the corresponding vacuum BH solutions.

\bigskip

This paper is organized as follows. In Section~\ref{sec_model} we present a general model, including $N$ complex scalar fields minimally coupled to gravity and the general ansatz for a solution with two different angular momenta. Various quantities of interest are described and the boundary conditions for the numerical implementation are presented. In particular, a near-horizon analysis immediately leads to condition (\ref{equilibrium}), from regularity. In section~\ref{sec_sing_am} we perform the analysis of single angular momentum solutions, starting with boson stars and addressing subsequently hairy BHs. Finally, in Section~\ref{sec_conclusions} we provide some final remarks.

\section{The general model}
\label{sec_model}

\subsection{Action and matter content}
 
We shall consider a model with $N$ complex scalar fields
$\Psi^{(I)}$ coupled to Einstein gravity in $D=5$,
\begin{eqnarray}
\label{action}
\mathcal{S}=\int  d^5x \sqrt{-g}\left ( \frac{1}{16\pi G}R
-\sum_{I=1}^N {\cal L}^{(I)} 
\right ) \ ,
\end{eqnarray}
where $G$, that will be set to unity, is Newton's constant and the Lagrangian density for each of the scalar fields is 
\begin{eqnarray}
\label{L-scalar}
{\cal L}^{(I)}= \frac{1}{2} g^{ab}
\left( 
\Psi_{,  a}^{(I)*} \Psi_{,  b}^{(I)} +  \Psi_{, b}^{(I)*} \Psi_{,   a}^{(I)}
\right)+U( |\Psi^{(I)}|) \ .
\end{eqnarray}
Thus, the scalar fields do not interact with one another. $U( |\Psi^{(I)}|)$ is the $I$-th scalar field potential. 
Variation of the action (\ref{action}) with respect to the metric  
 yields the Einstein  equations:
\begin{eqnarray}
\label{Einstein-eqs}
 R_{ab}-\frac{1}{2}g_{ab}R=8 \pi G~\sum_{I=1}^N T_{ab}^{(I)}\ ,
\qquad
{\rm where}
\qquad
T_{ab}^{(I)}= 
 \Psi_{ , a}^{(I)*}\Psi_{,b}^{(I)}
+\Psi_{,b}^{(I)*}\Psi_{,a}^{(I)} 
-g_{ab} {\cal L}^{(I)} \ ,
\end{eqnarray}  
is the
energy-momentum tensor  of the $I$-th scalar field.
 There are also $N$ Klein-Gordon equations, obtained by varying the action with respect to each of the scalar fields
 \begin{eqnarray}
\label{KG-eq}
 \nabla^2 \Psi^{(I)}= \frac{d U^{(I)} }{d  |\Psi^{(I)}|^2} \Psi^{(I)}\ .
 \end{eqnarray}

\subsection{The general ansatz}

To better understand the metric ansatz, split five dimensional Minkowski space as 
$\mathbb{M}^{1,4}=\mathbb{R}_t\times\mathbb{R}^2_{\varphi_1,\rho}\times \mathbb{R}^2_{\varphi_2,\sigma}$. 
Thus,  the four-dimensional Euclidean space is split into two 2-planes each parameterized with polar coordinates. 
The corresponding coordinate transformation between Cartesian and bi-polar coordinates in $\mathbb{R}^4$ is $x^1=\rho \sin\varphi_1$, 
$x^2=\rho \cos\varphi_1$,
$x^3=\sigma \sin\varphi_2$, 
$x^4=\sigma \cos\varphi_2$,
where $\rho,\sigma$
are polar radial coordinates
in the 2-planes,
$0\leq \rho,\sigma<\infty$ and $0\leq  \varphi_i<2 \pi$ are azimuthal angles. Rotations along $\varphi_1$ and $\varphi_2$ generate two independent angular momenta. 
The generic rotating solutions depend on both $\rho$ and $\sigma$;
however, the numerics and the description of solutions simplify by introducing a (hyper-)spherical radial coordinate in $\mathbb{R}^4$, $r$, and an angle $\theta$, such that the polar radii become projections of $r$ into each of the two 2-planes:
$\rho=r\sin\theta $, $\sigma=r\cos\theta $,
with 
$0\leq r<\infty$, $0\leq \theta\leq \pi/2$. Then $\partial_{\varphi_1}$ ($\partial_{\varphi_2}$) generates rotations in the plane $\theta=\pi/2$ ($\theta=0$) and $\mathbb{M}^{1,4}$ is written
\begin{equation}
ds^2_{\mathbb{M}^{1,4}}=-dt^2+ (dx^1)^2 + (dx^2)^2+ (dx^3)^2+ (dx^4)^2=-dt^2+dr^2+r^2(d\theta^2+\sin^2\theta d\varphi_1^2+ \cos^2\theta d\varphi_2^2) \ .
\label{flatm}
\end{equation}

The curved spacetimes we shall be considering contain corrections to the metric tensor (\ref{flatm}), which is only approached asymptotically. In general, we assume solely that the line element possesses
three commuting Killing vectors,
$\xi = \partial_t$, $\eta_1=\partial_{\varphi_1}$, and $\eta_2=\partial_{\varphi_2}$.
A suitable metric parametrization for a BH spacetime reads\footnote{A version of this ansatz 
has been employed
in the construction of $D=5$ counterparts 
of the Kerr-Newman solution \cite{Kunz:2005nm},
 generalizing the one used used in
\cite{Kleihaus:2000kg}
 to construct the first four dimensional
spinning hairy BHs in the literature.
}
\begin{eqnarray}
 \label{metric} 
&&ds^2 =  -F_0(r,\theta) N(r) dt^2+
 F_1(r,\theta)\left( \frac{dr^2}{N(r)}+r^2 d\theta^2 \right)
  +F_2(r,\theta) r^2 \sin^2\theta
          \left[ d\varphi_1-W_1(r,\theta) dt \right ]^2 ~~~~~~~~~~~~~~~~~~
\\
\nonumber 
&&~~~~~~+F_3(r,\theta) r^2 \cos^2\theta
          \big[ d\varphi_2-W_2(r,\theta) dt \big ]^2 +F_4(r,\theta)r^2 \sin^2\theta \cos^2\theta 
 \left[ W_2(r,\theta)  d\varphi_1 - W_1(r,\theta)  d\varphi_2\right]^2 \ ,
   \end{eqnarray}
   in terms of seven metric functions, $F_0,F_1,F_2,F_3,F_4,W_1,W_2$ and also
\begin{eqnarray}
\nonumber
N(r)\equiv 1-\frac{r_H^2}{r^2}\ ,
\end{eqnarray}
where the parameter $r_H\geq 0$ corresponds to the position of the BH horizon in this coordinate system.

The particular parametrization just described for the line element is compatible with an ansatz for the matter fields of the form:
\begin{eqnarray}
\label{scalar}
 \Psi^{(I)}= 
 \phi^{(I)}(r,\theta)e^{i\big(m_1^{(I)}\varphi_1+m_2^{(I)}\varphi_2-w^{(I)}t \big) } \ ,
 \end{eqnarray} 
where $m_i^{(I)}\in \mathbb{Z}$ are azimuthal harmonic indices in both planes of rotation 
and $w^{(I)}>0$ is the $I^{th}$ scalar field frequency. Observe that the three aformentioned Killing vector fields, $\xi = \partial_t$, $\eta_1=\partial_{\varphi_1}$, and $\eta_2=\partial_{\varphi_2}$, do not preserve, independently the scalar fields; rather, $\Psi^{(I)}$ are only preserved by the 2-parameter family of helicoidal Killing fields $\partial_t+\alpha_1\partial_{\varphi_1}+\alpha_2\partial_{\varphi_2}$ with
\begin{equation}
w^{(I)}=m_1^{(I)}\alpha_1+m_2^{(I)}\alpha_2 \ .
\label{helicoidal}
\end{equation}

\subsection{Global charges and other physical quantities}

We shall now present a set of physical quantities and relations that apply to the boson stars and the  MPBHsSH that shall be obtained in the next section.

The solutions approach Minkowski spacetime at infinity. Then, as usual, the ADM  mass $M$ and the ADM angular momenta $J_i$ can be read off from the asymptotics of particular metric functions,
\begin{eqnarray}
\label{asym}
g_{tt} =-1+\frac{8 M}{3\pi r^2}+\dots,
~~g_{\varphi_1 t}=-\frac{4 J_1}{\pi r^2}\sin^2\theta+\dots,
~~g_{\varphi_2 t}=-\frac{4 J_2}{\pi r^2}\cos^2\theta+\dots~.
\end{eqnarray}

For the   line element (\ref{metric}),
the event horizon $\mathcal{H}$ is a surface of constant radial coordinate,
$r=r_H$; $\mathcal{H}$ is a Killing horizon of the Killing vector field
\begin{equation}
\chi=\xi + \Omega_1 \eta_1 + \Omega_2 \eta_2 \ ,
\label{Killing} \end{equation}
which is null on $\mathcal{H}$ and orthogonal to it. Here, $\Omega_1=W_1\big|_{r_H}$ and $\Omega_2=W_2\big|_{r_H}$ denote the horizon angular velocities
with respect to rotation in the $\theta=\pi/2$ and $\theta=0$ plane,
respectively.  

MPBHsSH have Hawking temperature 
\begin{equation}
\label{TH}
T_H=\frac{1}{2\pi r_H}\sqrt{\frac{F_0}{F_1}}\bigg|_{r_H} \ , 
\end{equation}
and horizon area (related to the entropy by $S=A_H/4$)
\begin{equation}
~A_{ H} = (2 \pi)^2 r_{H}^3 
\int_0^{\pi/2}  d\theta \sin \theta  \cos \theta
 \sqrt{F_1\left(F_2F_3 +F_4(\sin^2\theta F_2 W_1^2+\cos^2\theta F_3 W_2^2)\right)} ~
\bigg |_{r_{ H}} \ .
\end{equation}

The Lagrangian of each scalar field  has a global $U(1)$ symmetry which introduces $N$
conserved currents $j^{a(I)}=-i (\Psi^{(I)*} \partial^a \Psi^{(I)}-\Psi^{(I)} \partial^a \Psi^{(I)*})$,
with $j^{a(I)}_{;a}=0$.
Thus the solutions carry also $N$ conserved Noether charges -- in the sense of obeying local continuity equations, but \textit{not} (global) Gauss laws -- obtained by integrating the Noether charge density, $j^t$, on a spacelike slice $\Sigma$,  
\begin{eqnarray}
\label{Q}
Q^{(I)}=\int_{\Sigma}dr d\theta d\varphi_1d\varphi_2  \sqrt{-g} ~j^{t(I)}\ .
\end{eqnarray}
MPBHsSH satisfy a Smarr-type relation
\begin{eqnarray}
\label{Smarr}
M=   M^{(\psi)}+ \frac{3}{2}
\left(
T_H S+\Omega_1 J_1+\Omega_2 J_2 
-\Omega_1 \sum_I m_1^{(I)}Q^{(I)} 
-\Omega_2 \sum_I m_2^{(I)}Q^{(I)} 
\right)\ ,
\end{eqnarray}
where $M^{(\psi)}$ measures  the energy stored in the scalar field outside the horizon:
\begin{eqnarray}
\label{Mpsi}
M^{(\psi)}=-\sum_{I}\frac{3}{2}\int _\Sigma d^4x \sqrt{-g}
\left(
T_t^{(I)t}-\frac{1}{3}T_a^{(I)a}  
\right) \ .
\end{eqnarray}

A Smarr-type relation involving only horizon quantities also exists
\begin{eqnarray}
\frac{2}{3} M_{ H}=  T_H S
+ \Omega_1 J_{1,  H} + \Omega_2 J_{2,  H}
\ , 
\end{eqnarray}
with
\begin{eqnarray}
J_{i,  H}=J_i-\sum_I m_i^{(I)}Q^{(I)} \ .
 \end{eqnarray}

Finally, MPBHsSH satisfy the first law of thermodynamics
\begin{eqnarray}
\label{1st}
dM=T_H dS+\Omega_1 dJ_1+\Omega_2 dJ_2\ .
\end{eqnarray}

\subsection{Boundary conditions}

As for the case of Kerr BHs with scalar hair~\cite{Herdeiro:2014goa}, 
and for the case of MPBHsSH with two equal angular momenta~\cite{Brihaye:2014nba}, 
there are no exact solutions in closed form of the above system with a non-trivial scalar field. 
The problem can, however, be tackled numerically,
by solving a set of elliptic equations with given boundary conditions.
 
To obtain asymptotically flat solutions with finite mass, we impose
 the boundary conditions  at infinity  
\begin{equation}
 \label{bc-inf} 
F_0=F_1=F_2=F_3=F_4=1,~~W_1=W_2=0,~~~\phi^{(I)}=0 \ .
\end{equation}

At $\theta=0,\pi/2$ the metric functions satisfy Neumann boundary conditions.
The boundary conditions for the scalar   field amplitude $\phi^{(I)}$
are more complicated.
In the generic case with $m_1^{(I)}\neq 0$,  $m_2^{(I)}\neq 0$,
 $\phi^{(I)}$ vanishes at $\theta=0,\pi/2$.
However, for $m_1^{(I)}\neq 0,~m_2^{(I)}= 0$, the scalar field amplitude $\phi^{(I)}$
vanishes at $\theta=0$ only, and  
satisfies Neumann boundary condition
at $\theta=\pi/2$; the case $m_1{(I)}= 0,~m_2^{(I)}\neq 0$ 
follows immediately, \textit{mutatis mutandis}.  

The boundary conditions on the horizon take a simpler form in terms of a  new radial variable $x=\sqrt{r^2-r_H^2}$ (which is also employed in numerics): 
$\partial_x F_i\big|_{\mathcal{H}}=\partial_x \phi^{(I)}\big|_{\mathcal{H}}=0$.
Of central importance, regularity at the horizon implies that the following \textit{resonance condition}
should be satisfied for each scalar field
\begin{eqnarray}
\label{resonance}
w^{(I)}=m_1^{(I)}\Omega_1+m_2^{(I)}\Omega_2  \ .
 \end{eqnarray}
Comparing with (\ref{helicoidal}) this condition singles out a particular helicoidal Killing vector field within the family that preserves the full ansatz (\ref{metric}) plus (\ref{scalar}), corresponding to $\alpha_i=\Omega_i$, in other words, the one that coincides with the BH horizon generator. 

Finally, the metric functions should satisfy the
  elementary flatness conditions, guaranteeing absence of conical singularities on the axis:
\begin{equation}
(F_2+F_4 W_2^2-F_1=0)|_{\theta=0},~~~(F_3+F_4 W_1^2-F_1=0)|_{\theta=\pi/2}.
\end{equation}

\section{Single angular momentum solutions}
\label{sec_sing_am}
We shall now specify the general ansatz (\ref{metric})--(\ref{scalar})  and the general model (\ref{action}), by focusing on the following special case:
\begin{description}
\item[i)] We consider a single ($N=1$) massive but non-self-interacting scalar field, such that $U( |\Psi^{(I)}|)= \mu^2 |\Psi^{(I)}|^2$, where $\mu$ is the scalar field mass and $I=1$. Thus, from now on we shall drop the superscript $I=1$, as there will only be a single complex scalar field.
\item[ii)] We focus on solutions with rotation on a single plane. Then, one can set 
\begin{equation}
m_1\neq 0,~~m_2=0
\end{equation}
in the scalar field ansatz (\ref{scalar}), which in particular implies that $T_{\varphi_2}^t=0$, and thus one can consistently set $F_4=W_2=0$
in the line-element (\ref{metric}). For simplicity of notation, in the following we shall drop the subscript $1$ referring to the plane of rotation ($e.g.$ $m_1\rightarrow m$).
\end{description}

\subsection{The vacuum limit: Myers-Perry BHs}
\label{sec_mp1}
Setting $\phi=0$ in (\ref{scalar}), the model described in Section~\ref{sec_model} admits as solutions MPBHs~\cite{Myers:1986un}, which are exact solutions known in closed form. MPBHs with a singular angular momentum parameter (in $D=5$)  can be written in the form of our ansatz (\ref{metric}), with:
\begin{eqnarray}
\nonumber
&&
F_1(r,\theta)=1+\frac{a^2}{r^2}\cos^2 \theta \ ,~~~~F_2(r,\theta)=\left(1+\frac{a^2}{r^2}\right)\left(
1+\frac{a^2 \sin^2\theta}{\frac{(a^2+r^2)(r^2+a^2\cos^2\theta)}{(a^2+r_H^2)}}
\right) \ ,
~~~~F_3(r,\theta)=1\ ,
\\
\label{MP-metric}
&&
F_0(r,\theta)=\frac{1}{F_2}\ ,~~W_1(r,\theta)=\frac{1}{F_2}\frac{a(r_H^2+a^2)}{r^2(r^2+a^2\cos^2\theta)}\ , 
\end{eqnarray}
which apart from $r_H$, contains the extra parameter $a$ associated with rotation. 
Some physical quantities are given, in terms of the parameters $r_H,a$ as
\begin{eqnarray}
\label{MP-quant}
\nonumber
M=\frac{3\pi }{4 }(a^2+r_H^2)  ,~ 
J=\frac{\pi }{2 }a(a^2+r_H^2)   ,~
T_H=\frac{r_H}{2\pi(a^2+r_H^2)},~
A_H=4 \pi^2 r_H(a^2+r_H^2)  ,~\Omega_H=\frac{a}{a^2+r_H^2}.
\end{eqnarray}
%
The properties of these solutions have been extensively discussed in the literature.
Here we mention only that the spinning BHs are continuously connected to the Schwarzschild-Tangherlini solution in the static limit; also, in contrast to the $D=4$ Kerr metric, 
the zero temperature limit (which corresponds to $r_H\to 0$ for nonzero $a$) 
is singular in this case, with $A_H\to 0$ \cite{Bardeen:1999px}.


\subsection{The solitonic limit: Boson Stars}
Turning on the scalar field, we have found both solitonic (\textit{boson stars}) and BH solutions. These cannot, however, be found in closed form and we have resorted to numerical methods. The numerical approach employed here is similar to that used in constructing $D = 4$ Kerr BHs with  scalar hair described in  \cite{Herdeiro:2014goa}. 
As usual, dimensionless variables and global quantities are introduced by using
natural units set by $\mu$ (we recall $G=1$), $e.g.$ $r\to r \mu$,
 $\phi \to \phi  /\sqrt{8\pi}$ and $w \to w/\mu$. 
Then,  the numerical treatment of the model relies on only four input parameters:   
the horizon radius $r_H$ (for BHs), 
the field frequency $w$,
the winding number $m$
and the scalar field node number $n$. 
In the following we shall only consider nodeless solutions $(n=0)$ corresponding to the fundamental state of boson stars and hairy BHs. 

The equations for the $F_0,F_1,F_2,F_3,W,\phi$   are solved by using
a professional finite difference solver \cite{schoen},
which provides an error estimate for
each unknown function.
Other numerical tests were provided by the Smarr relation
(\ref{Smarr}) and the first law (\ref{1st}).
Based on that,  
the typical numerical error for the solutions here is estimated
to be around $10^{-3}$.

\begin{figure}[h!]
\centering
\includegraphics[height=2.9in]{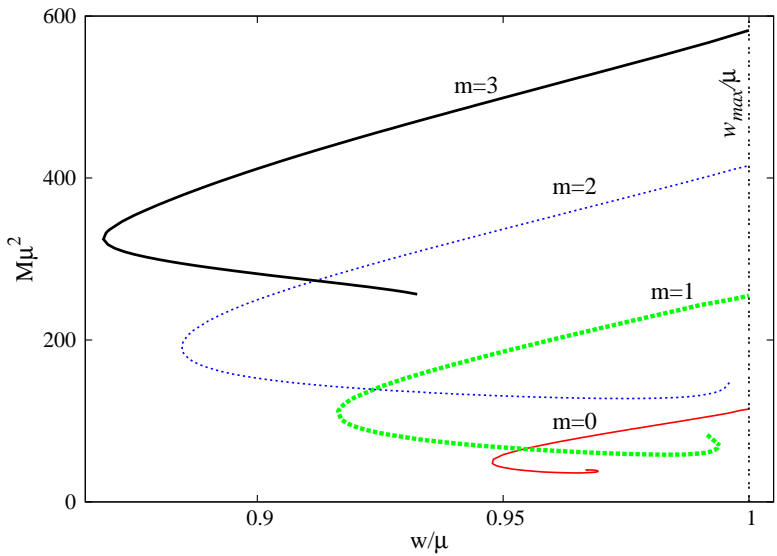}\\
\caption{
{\small
ADM mass $vs.$ scalar field frequency diagram for boson stars, with $m=0,1,2,3$.
}
} 
\label{BS}
\end{figure}

Setting $r_H=0$ in (\ref{metric})  the horizon is replaced with a regular origin and one finds boson star solutions. 
Up to now, only co-dimension one problems have been studied:
 boson star solutions have been reported both within spherical symmetry~\cite{Astefanesei:2003qy}, 
 and with two equal angular momenta \cite{Hartmann:2010pm}; the latter are, however found for a model with two complex scalar fields. The boundary conditions at the origin are  
 similar to those described above, except for the metric function $W$, which satisfies now a Neumann boundary condition $\partial_r W=0$. 

The Noether charge and the angular momenta of these boson stars are not independent quantities; they are simply related by 
\begin{eqnarray}
J=m Q\ ,
\end{eqnarray}
 while the Smarr relation and the first law read  
\begin{eqnarray}
 M=M^{(\psi)},~~dM= m w dJ \ .
\end{eqnarray} 

Taking the scalar field frequency $w$ as a control parameter, 
the numerical results show that, for any $m$, boson stars exist for a limited range of frequencies,
$w_{min}  <w< \mu$,
with $w_{min}(m)$
decreasing with $m$ -- Figure \ref{BS}. A striking
property of the $D=5$ boson stars is that these do not possess a true vacuum limit.
That is,
in contrast to the $D=5$ Anti-de-Sitter case~\cite{Dias:2011at}, or to the case of $D=4$ spinning boson stars~\cite{Yoshida:1997qf,Kleihaus:2005me}, $D=5$ asymptotically flat solutions do not trivialise as
$w\to \mu$. Indeed, as  noticed  in~\cite{Hartmann:2010pm}
for the special case of $D=5$ boson stars with two equal angular momenta,
as the frequency tends to the upper bound set by $\mu$,
  the scalar field spreads and tends to zero
while the geometry becomes arbitrarily close to
the Minkowski one.  The global charges of the solutions, however, 
remain finite and nonzero as  $w \to \mu$. Thus a mass (and charge) gap is found between the $\phi=0$
 vacuum  flat space ground state and the limiting configurations with
a frequency $w$ arbitrarily close to $\mu$. 
This behaviour has been explained for spherically symmetric solutions 
and for the two equal angular momenta boson stars~\cite{Hartmann:2010pm}, observing the existence 
of a special scaling symmetry of the limiting solutions. It seems plausible 
that the results in \cite{Hartmann:2010pm} can be extended to the case of boson stars with a single angular momentum.

The results of the numerical integration for several values of $m$
are displayed in Figure \ref{BS}. For completeness, we have included there also the  case of
spherically symmetric boson stars,  which can also be studied within the general ansatz 
(\ref{metric})--(\ref{scalar}), by taking $m_i=0$, $F_1=F_2=F_3$, $W_i=0$, 
and the surviving three independent functions, $F_0$, $F_1$ and $\phi$ depending only on $r$ 
(note also that in this case the scalar field does not vanish at $r=0$).

From Figure \ref{BS}
we observe that the mass $M$ decreases as $w$ is decreased from the maximal value $\mu$.
After approaching the minimal value $w_{min}$, a
backbending in $w$ is observed.
Then, one expects an inspiralling behaviour of the curves, towards a limiting configuration
at the center of the spiral, for a frequency $w_{cr}/\mu$. This part of the diagram is difficult
 to explore numerically for spinning solutions,   and so a second backbending is only clearly shown for $m=0,1$. 
 This inspiraling pattern 
 appears to be generic for boson star solutions\footnote{This part of the diagram
 appears to 
 change, however, for solutions with two equal angular momenta 
 in the $D=5$ Einstein-Gauss-Bonnet model
 \cite{Brihaye:2013zha,Henderson:2014dwa}.}, being found also  for boson stars in $D=4$ Einstein gravity
and a scalar-tensor extension \cite{Kleihaus:2015iea}, for  $D=5$
solutions with Anti-de-Sitter asymptotics \cite{Dias:2011at}
and for $D=5$
asymptotically flat solutions~\cite{Brihaye:2014nba}.
A similar diagram is found for $J(w)$, showing that boson stars do not possess a slowly rotating limit.

\begin{figure}[h!]
\centering
\includegraphics[height=2.50in]{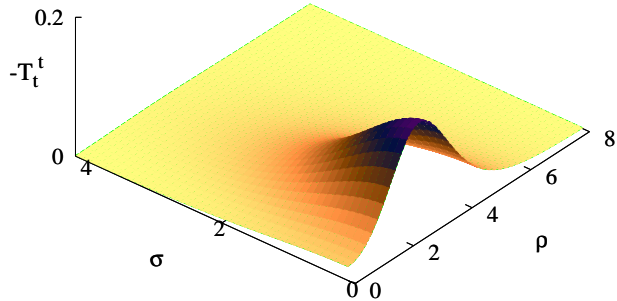}\\
\caption{
{\small
The $T_t^t$-component of the energy momentum tensor is shown  as a function of 
the coordinates $\rho=r \sin \theta$, $\sigma=\rho \cos\theta$, for a nodeless boson star  with $m=1$, $w/\mu=0.93$. Here, $T_t^t$,  $\rho$ and $\sigma$ are given in units of the scalar field mass $\mu$.}
}
\label{BS-rho}
\end{figure} 

 The $T_t^t$-component of the energy-momentum tensor of a typical boson star is shown in Figure \ref{BS-rho}.
 There one can notice the existence of a maximum in the plane of rotation, for some nonzero value of $r$. 

\subsection{Hairy black holes}

In order to obtain MPBHsSH we consider $r_H\neq 0$. Turning on this parameter, starting from any
given boson star solution with frequency $w$, can be regarded as adding a small BH at the center 
of the boson star. For a given $\Omega_H$, the boson star with  $r_H=0$
provides a good initial profile for hairy BHs with a small $r_H$. 
By increasing $r_H$ from zero, we obtain MPBHsSH with $\Omega_H$
fixed by the scalar field frequency.
It follows that the minimal frequency of the boson stars sets a lower bound on the horizon velocity of the hairy BHs, while the upper bound on the  frequency is still set by $\mu$,
the scalar field mass.

Given this systematic construction technique, it is convenient to describe the domain of existence of the hairy BHs in terms of $\Omega_H$.
The emerging picture
shows that, when varying the horizon size (via
the parameter $r_H$),  
 there are two 
possible types of sequences of BH solutions with a fixed $\Omega_H$:

\bigskip

$(S1)$   There are sequences of BH solutions that connect two different boson star solutions with the same frequency. Along these sequences, the BH solutions attain a maximal area at some point in between the two boson star solutions with the same scalar field frequency. Approaching these solutions, $r_H\to 0$, the horizon area vanishes, the temperature
 diverges and $J \to mQ$. For $m=1$, this occurs, for instance, for frequencies between the minimal boson star frequency $w_{min}$ and $w/\mu\simeq 0.936$.

\bigskip

$(S2)$
 There are sequences of BH solutions that end in a zero temperature 
 extremal BH with scalar hair.
 In contrast to both Kerr BHs with scalar hair~\cite{Herdeiro:2014goa} and the MPBHsSH studied in~\cite{Brihaye:2014nba}, these limiting configurations have vanishing
horizon size and do not seem to possess 
a regular horizon. The global charges, however, are finite and nonzero in this case.

\bigskip

\begin{figure}[h!]
\centering
\includegraphics[height=2.48in]{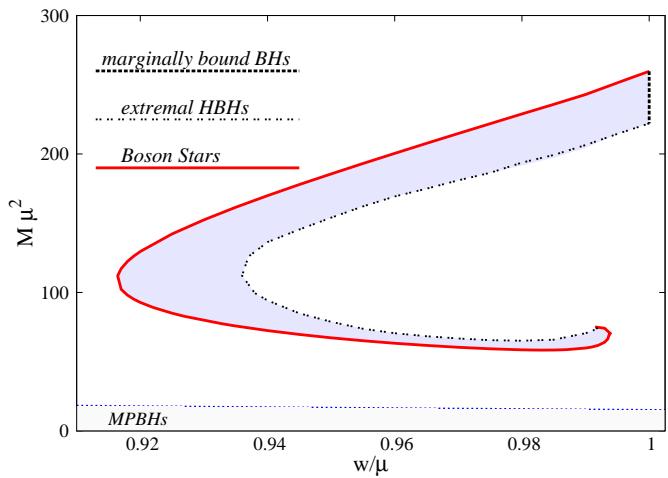}\\
\caption{ 
{\small
The (extrapolated) domain of existence  of MPBHsSH (shaded blue region) with an azimuthal harmonic index $m=1$ in $(M,w)$-space.
}
} 
\label{Mw}
\end{figure}

In Figure \ref{Mw}  we show the domain of existence of MPBHsSH
(the shaded blue region), for solutions with $n = 0,~m = 1$,
as a function of frequency $w$. This domain was obtained by extrapolating to the continuum
the results from a set of around two thousand  numerical solutions.
 This can safely be done
for most of the parameter space. We cannot exclude, however, a more complicated picture for a small region
around the 
center of the boson star spiral, which is rather difficult to explore numerically
within our approach. 
We 
further remark that the set  of extremal MPBHsSH which form a part of the boundary of the domain of existence have been obtained by extrapolation of the numerical results\footnote{Differently from the hairy BHs in 
\cite{Herdeiro:2014goa,Brihaye:2014nba}, the direct construction of this set of extremal BHs presented unsurmountable difficulties, presumbably due to their singular nature. Indeed, we observed that 
for the near extremal solutions,  both the Ricci and the Kretschmann scalars
take very large values on the horizon, in particular at $\theta=\pi/2$.
}.
 
The domain of existence presented in Figure \ref{Mw} is delimited by three curves: the already discussed boson star curve (red solid line), the curve of \textit{extremal} ($i.e.$ zero temperature)  MPBHsSH (black dashed line), and a vertical line segment with $w=\mu$ which correspond to the limiting configurations dubbed \textit{marginally bound solutions} (black dotted line)~\cite{Brihaye:2014nba}. We remark that a similar diagram is found for $J(w)$. 
Thus, we conclude that MPBHsSH with a singular angular momentum have a minimal mass and angular momentum. In particular they have no static limit, analogously to Kerr BHs with scalar hair \cite{Herdeiro:2014goa}. Figure \ref{Mw} focuses on $m=1$;  based on preliminary numerical data, we are confident 
that a similar pattern for the domain of existence of MPBHsSH occurs for other values of $m$.

 The line describing the extremal
 solutions 
 starts at a non-zero ADM mass at the maximal frequency $w/\mu\to 1$, 
 decreases until a minimal value of 
 $w/\mu$ (with $w/\mu\simeq 0.936$ for $m=1$),
  backbends and keeps decreasing, reaches a minimal value of the ADM mass  and then seems to inspiral towards a central value where, we conjecture,  
 it meets the endpoint of the boson star spiral in a singular solution.
  
  \bigskip
  
\begin{figure}[h!]
\centering
\includegraphics[height=2.2in]{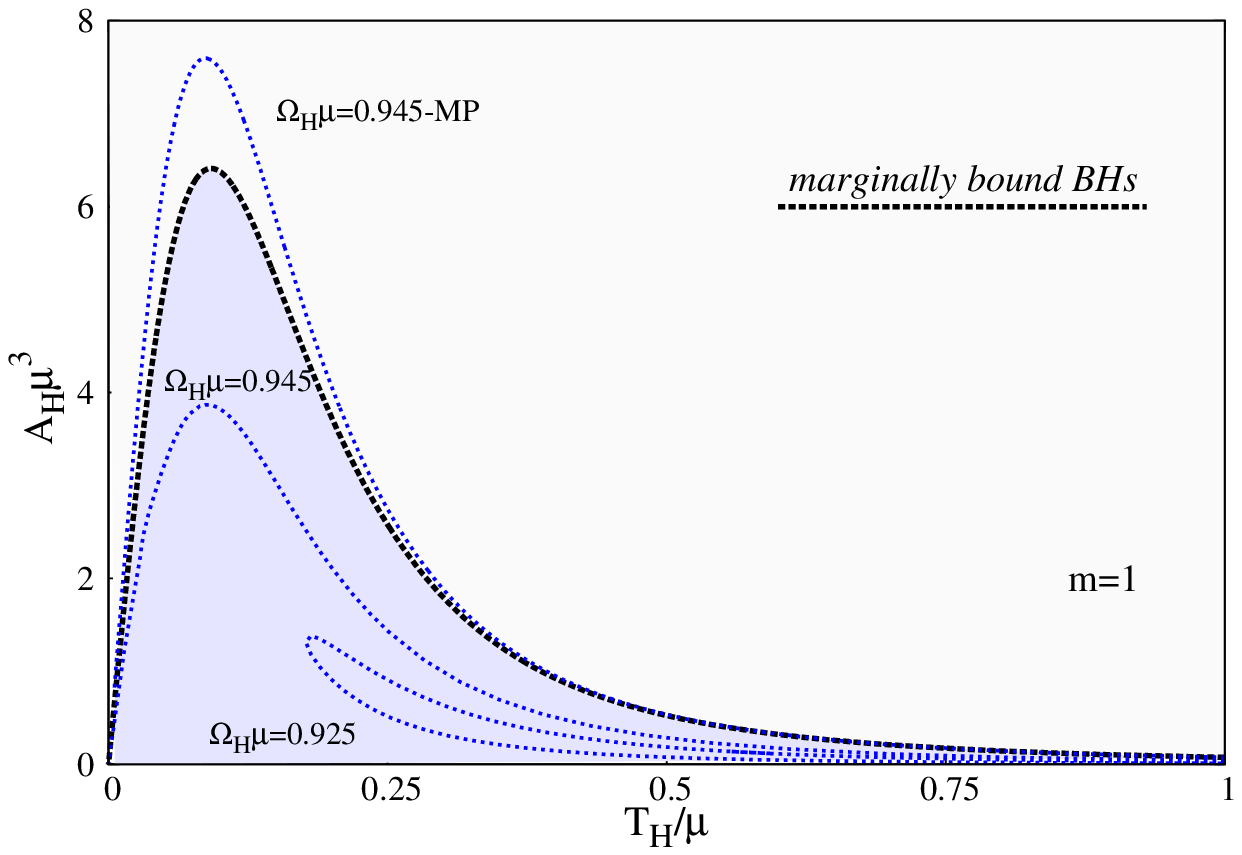}
\includegraphics[height=2.2in]{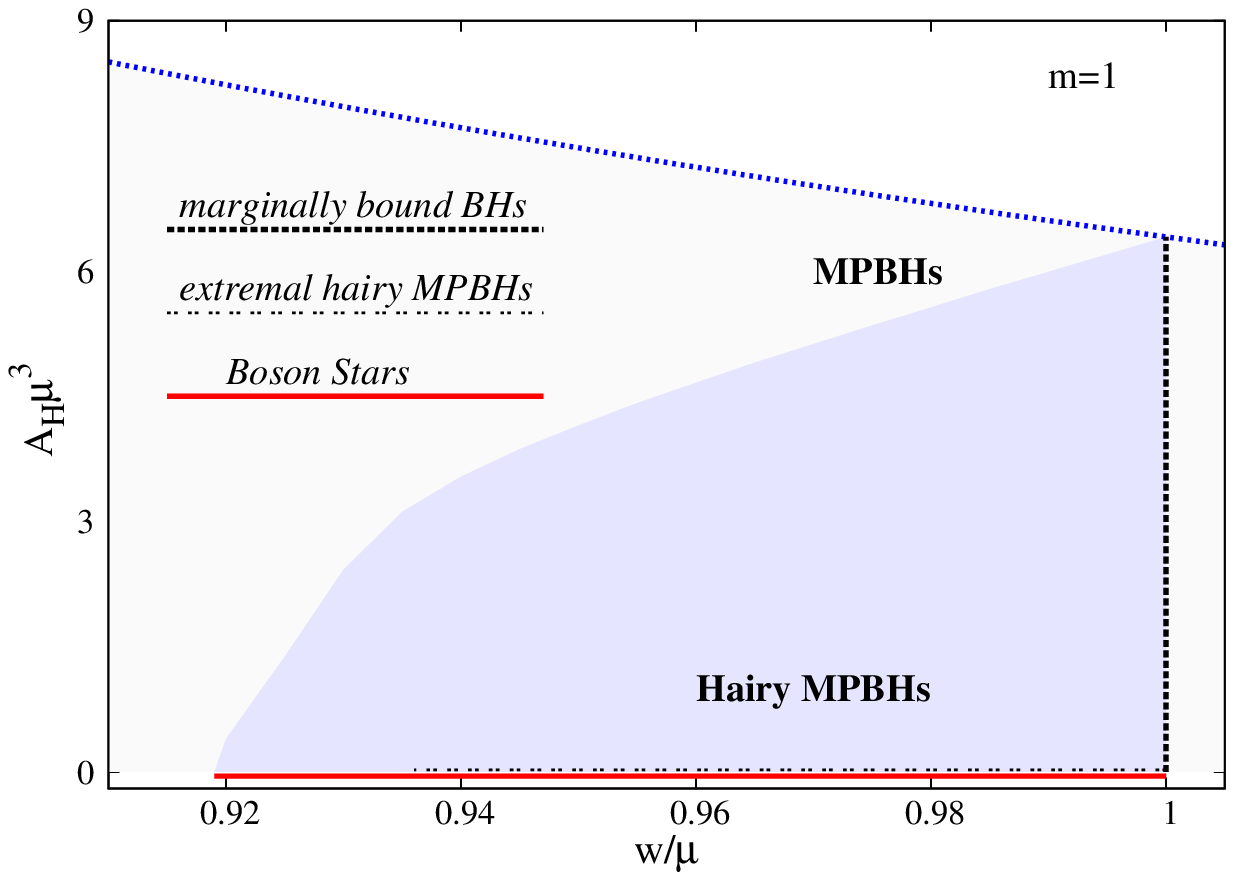}
\caption{
{\small
{\it (Left panel)} 
Domain of existence  of MPBHsSH (blue shaded area) in a horizon area $A_H$ \textit{vs} temperature diagram. 
We have also plotted lines of constant horizon angular velocity $\Omega_H$
 (blue dotted curves). When $\Omega_H=\mu$ the line of vacuum MPBHs precisely coincides with that of the marginally bound MPBHsSH.  
{\it (Right panel)}
 Domain of existence in a horizon area \textit{vs} frequency (equal to $\Omega_H$) diagram. Vacuum MPBHs exist below the blue dotted line. For $\Omega_H=\mu$, their domain of existence coincides precisely with the marginally bound MPBHsSH.
}
}
 
\label{areatemperature}
\end{figure}

  Further features of singly spinning MPBHsSH are shown in Figure \ref{areatemperature},
  where we plot 
  their domain of existence in a horizon area $A_H$ $vs.$
  temperature diagram (left panel), 
 and in a $A_H$  $vs.$ frequency diagram (right panel). The results for vacuum MPBHs are also shown for comparison.
As one can observe from the left panel, for a given frequency, the horizon area reaches 
 a maximal value for some solution with nonzero $T_H$. Let us consider two qualitatively distinct examples. For $\Omega_H/\mu=0.945$ the sequence of solutions interpolates between infinite temperature (a boson star) and zero temperature (an extremal MPBHSH), corresponding to a sequence of type S2 above. By contrast, for $\Omega_H/\mu=0.925$ the sequence interpolates between two boson stars (hence two infinite temperatures), corresponding to a sequence of type S1 above. Note that for $\Omega_H/\mu=0.945$ we have also plotted a sequence of vacuum MPBHs.  In the $A_H$ $vs.$ $w$ diagram,
 the set of critical configurations with maximal area for fixed $\Omega_H$ form a part of the boundary
 of the domain of existence\footnote{This is the behavior found also for MPBHs.
For a given angular velocity $\Omega_H$, the horizon area of a MPBH approaches 
 a maximal value for $a=3/(4\Omega_H)$ where $T_H=\frac{\Omega_H}{2\pi \sqrt{3}}$.
The horizon area decreases for larger values of $a$, and approaches zero
for the maximal value $a\to \Omega_H$ which corresponds to an extremal (singular) configuration.
 }.
 The remaining boundary is given by the set of boson stars, which have $A_H=0$, together with the extremal MPBHsSH, which have also zero horizon area and the set of marginally bound BHs.
 
\begin{figure}[h!]
\centering
\includegraphics[height=2.48in]{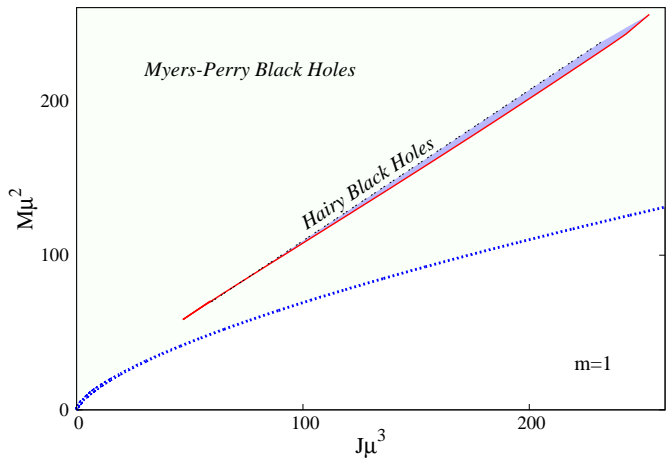}
\\
\caption{ {\small Extrapolated domain of existence (very thin shaded area) of
the $m=1$ MPBHsSH in the $(J,M)$ space. Vacuum MPBHs exist above the blue dotted line, thus overlapping with the former solutions. The colour coding here is similar to that used in Figure 4.
} 
}
\label{jm}
\end{figure}

From Figure~\ref{areatemperature} it can also be observed that there is continuity between vacuum MPBHs and MPBHsSH in terms of \textit{horizon quantities}, as was observed in~\cite{Brihaye:2014nba} for the two equal angular momentum case. This occurs despite the mass gap between the two families of solutions in terms of global charges.  

Finally, in Figure \ref{jm} we 
plot the phase space of MPBHsSH, \textit{i.e.} the domain of existence of these BHs 
 in the $(J,M)$-plane. As it can be observed they exist in the region where vacuum MPBHs exist as well. As such there is non-uniqueness, when only the ADM mass and angular momentum are specified, in analogy to the case of Kerr BHs with scalar hair~\cite{Herdeiro:2014goa}.

\section{Further remarks}
\label{sec_conclusions}
In this paper we have reported the first construction of higher dimensional ($D>4$) boson stars and scalar hairy BHs with a single angular momentum parameter in the literature. One of the conclusions of our study is the confirmation that the properties of scalar hairy BHs within this large family of solutions anchored on conditions of type (\ref{equilibrium}) are inherited from their ``building blocks", which in the case considered herein are $D=5$ singly spinning boson stars and Myers-Perry BHs. Thus, MPBHsSH have a mass gap with respect to the vacuum MPBHs, as do boson stars with respect to Minkowski spacetime. Moreover, in the extremal limit, MPBHsSH yield a singular configuration as vacuum MPBHs do. This reinforces the picture that such hairy BHs can be viewed as bound states of ``bald" BHs and solitonic configurations (boson stars)~\cite{Herdeiro:2014ima}.

\bigskip

We have not considered here the general case with two non-vanishing angular momenta. 
In Ref.  \cite{Brihaye:2014nba}, however, MPBHsSH with  two equal angular momenta were studied 
in a model with  $N=2$ complex scalar fields.
Therein a special ansatz is used, originally proposed in \cite{Hartmann:2010pm},
such that the spacetime isometry group is enhanced  from $\mathbb{R}_t \times U(1)^{2}$
to $\mathbb{R}_t \times U(2)$.
This  enhancement is obtained by taking the same mass and frequency for both complex scalars, and requiring the fields to  rotate with the \textit{lowest} azimuthal harmonic index in different planes:
\begin{equation}
w^{(1)}=w^{(2)}=w,~~\mu^{(1)}=\mu^{(2)}=\mu,~~{\rm and}~~m_1^{(1)}=m_2^{(2)}=1,~~m_1^{(2)}=m_2^{(1)}=0\ ,
 \end{equation}
 such that the resonance condition
 (\ref{resonance})
is fulfilled by each scalar.
Then the $\theta-$dependence factorizes
\begin{eqnarray}
\phi^{(1)}=\phi(r)\sin\theta \  ,~~
\phi^{(2)}=\phi(r)\cos\theta \ ,~~
 \end{eqnarray} 
while
the metric functions $F_i,W$ in (\ref{metric}) depend only on $r$, with 
$W_1=W_2=W(r),~F_4(r)=F_3(r)-F_2(r)$,
and the problem is \textit{effectively} co-dimension one\footnote{An explanation of this fact is given in Ref.
\cite{Stotyn:2011ns},
together with a generalization of the ansatz to higher odd dimensions.}.
The general properties of these MPBHsSH with $J_1=J_2$ are similar to those found
in this work for MPBHsSH  with a single $J$.
The main difference concerns the extremal solutions,
which, therein -- and similarly to the behaviour of vacuum MPBHs with two equal angular momenta --
have finite (and nonzero) horizon size and global charges and possess a regular horizon.

\bigskip
Based on the results in this paper and those in~\cite{Brihaye:2014nba} one can make an educated guess for  
the general case with two non-vanishing and non-equal angular momenta.
The domain of existence of such MPBHsSH will be bounded by the corresponding boson stars, by a set of marginally bound solutions -- which have a mass gap with respect to the vacuum MPBHs -- and the extremal limit will have the same properties as those of the corresponding vacuum MPBHs. 
A different state of affairs, however, will certainly be found in the asymptotically Anti-de-Sitter case. 
Singly spinning MPBHs are afflicted by the superradiant instability 
of a massive scalar field and thus singly spinning MPBHsSH continuously connected 
to MPBHs in Anti-de-Sitter should also exist, 
similarly to the equal angular momentum case~\cite{Dias:2011at}.

\bigskip

Finally we remark on two further possible generalizations. Firstly, as it is well known, $D=5$ vacuum gravity admits other solutions with different horizon topologies, most notably black rings~\cite{Emparan:2001wn}. It seems plausible that black rings with scalar hair anchored to the condition (\ref{equilibrium}) also exist, even if finding them numerically may be challenging. Secondly, going to $D>5$, MPBHs exhibit yet a qualitatively new feature: the existence of ultra-spinning BHs. It would certainly be interesting to construct both singly spinning boson stars and singly spinning MPBHsSH in $D>5$ to see if/how this new possibility impacts on such solutions.

\vspace{0.5cm}

\section*{Acknowledgements}
J.K. would like to acknowledge support by the DFG Research Training Group
1620 ``Models of Gravity'' and
by FP7, Marie Curie Actions, People, International Research Staff Exchange Scheme (IRSES-605096).
The work of C.H. and 
E.R. has been supported by the grants PTDC/FIS/116625/2010,  
NRHEP--295189-FP7-PEOPLE-2011-IRSES and by the CIDMA strategic funding UID/MAT/04106/2013.
B.S.
 acknowledges 
partial support from DIKTI research grant No. 00324.81/IT2.11/PN.08/2015.

 \begin{small}
 
 \end{small}

\end{document}